\documentstyle[prl,aps,floats,twocolumn,graphics,epsf]{revtex}
\begin{document}
\newcommand{\beq}{\begin{equation}}
\newcommand{\eeq}{\end{equation}}
\newcommand{\beqn}{\begin{eqnarray}}
\newcommand{\eeqn}{\end{eqnarray}}
\newcommand{\dpf}{\displaystyle\frac}
\newcommand{\no}{\nonumber}
\newcommand{\ep}{\epsilon}
\twocolumn[\hsize\textwidth\columnwidth\hsize\csname @twocolumnfalse\endcsname
\title
{Entropy and holography constraints for inhomogeneous universes}
\author
{Bin Wang$^{a,b}$, Elcio Abdalla$^{a}$ and Takeshi Osada$^{a}$}
\address
{$^{a}$ Instituto de F\'{\i}sica, Universidade de S\~{a}o Paulo,\\
C.P.66.318, CEP 05315-970, S\~{a}o Paulo, Brazil \\
$^{b}$ Department of Physics, Shanghai Teachers' University,
P.R. China}
\maketitle

\begin{abstract}
We calculated the entropy of a class of inhomogeneous dust
universes. Allowing spherical  symmetry, we proposed a holographic 
principle by reflecting all physical freedoms on the  
surface of the apparent horizon. In contrast to flat homogeneous
counterparts, the principle may break down in some models, 
though these models are not quite realistic.  
We refined fractal parabolic solutions to have a reasonable entropy value 
for the present observable universe and found that the holographic principle
always holds in the realistic cases.\\

PACS number(s): 98.80.Hw, 04.20.-q,  04.60.-m, 95.75.-z

\end{abstract}
\hspace{.2in}

]
\vspace{6ex} \hspace*{0mm} 
In view of the example of black hole entropy \cite{bekenstein}, an influential 
holographic principle relating the maximum number of degrees of freedom in
a volume to its boundary surface area has been put forward 
recently \cite{Hoof93,Sus95}. This principle is  
viewed as a real conceptual change in our thinking about gravity. The main
aim of the holographic principle is to generalize its application 
to a broader class of situations, 
including cosmology. However, in a general cosmological setting, there is
no unique appropriate notion 
which is analogous to the event horizon in black hole serving as a natural
boundary. This makes the generalization particularly difficult. 
A remarkable progress has been made by 
Fischler and Susskind (FS) \cite{FS98}. They have shown that for flat and
open Friedman-Lema\^{\i}tre-Robinson-Walker (FLRW) universes the
holographic principle holds with the total entropy of the matter inside 
the particle horizon being smaller than the area 
of the horizon. Various different modifications of FS version of the
holographic principle have been raised subsequently 
\cite{many}--\cite{Ven99}. Motivated by the fact that the first sucessful  
implication of AdS/CFT duality to solve the problem of the microscopic
interpretation of black hole entropy appeared in (2+1)-dimensional models
\cite{Stro98}, we formulated the holography in such a case 
\cite{W-A99,W-A00}. Recently Bousso has provided a more 
elegant and a broader holographic principle and has applied to a number of
examples 
including the recollapsing FLRW cosmological models \cite{Bou99}. Part of
Bousso's 
proposal has been proved in \cite{FMW99}, however there is still some
difficulties 
associated with it \cite{Lowe99}. 

It is of great interest to take a closer look of holography in a generic
realistic  
inhomogeneous cosmological setting. The first attempt was carried out by
Tavakol and Ellis  
\cite{TE99} who considering Bousso's proposal as well as a modified
version of it.  
In both cases they found that operational difficulties exist in
constructing the holographic  
principle in a realistic universe. 

Compared to the homogeneous universe, there is a further difficulty in
setting up the 
holographic principle in the real cosmos. In the homogeneous universes the
comoving 
entropy density is assumed to be a constant and the total entropy is just
the entropy 
density times the comoving volume. In addition to the vector defined to
describe the 
gravitational entropy flux \cite{Bonn85} and evolution of the density
contrast studied to 
answer the possible existence of gravitational entropy \cite{MT99}, until
now there is no 
exact calculation of the entropy of inhomogeneous universes. 

In the present paper, we concentrate our attention on the parabolic
Lemaitre-Tolman-Bondi (LTB) model which is the natural generalization of
the flat dust 
FLRW model. Considering some characteristics of the realistic models, such
as spherical 
symmetry, not referring to either initial or final moment, we find that it
is appropriate 
to adopt the idea suggested in \cite{BR99} for homogeneous universes by
defining the 
apparent horizon as a boundary hypersurface to construct the holographic 
principle here.
We will show that choosing the apparent horizon in the formulation of the
cosmic holography 
for the real cosmos is simple and valuable. From the first law of
thermodynamics we will 
define the entropy density in the inhomogeneous universe and calculate the
total entropy 
value within the apparent horizon. Different from the homogeneous flat
universe, we will 
show that in the general parabolic LTB models, some fractal universe will
violate the 
holographic principle. In order to describe the real cosmos, we will
refine fractal 
parabolic solutions by comparing the calculated entropy in fractal models
to that in our 
present observable universe. In the refined realistic models holographic
principle can 
always be satisfied. 

In normalized comoving coordinates the metric of the parabolic LTB model is
\begin{eqnarray}     
ds^2&=&-dt^2+R'^2dr^2+R^2(d\theta^2+\sin^2\theta d\phi^2)\nonumber\\
&=&h_{ab}dx^adx^b+\tilde{r}^2(x)(d\theta^2+\sin^2\theta d\phi^2) \label{1}
\end{eqnarray}
where $h_{ab}=diag \lbrack -1, R^{\prime 2}\rbrack $, and
\beqn       \label{2}
R & = & \dpf{1}{2}(9F)^{1/3}(t+\beta)^{2/3}        
\eeqn
is the area distance and $R'$ plays the role of scale factor. $\beta$
and $F$ are two 
arbitrary functions of $r$. 
The metric (\ref{1}) is spherically symmetric. We define the dynamical apparent
horizon in terms of a condition $\vert\vert\nabla \tilde{r}\vert
\vert^2\equiv h^{ab}\partial_a\tilde{r}\partial_b\tilde{r}=0$ to the 
areal radius, with the result 
\beq \label{4}
F(r_{AH})=3[t+\beta(r_{AH})] 
\eeq
and $ \tilde{r}_{AH}=\dpf{3}{2}[t+\beta(r_{AH})] $ is the physical 
apparent horizon, where $r_{AH}$ denotes the proper apparent horizon. Taking  
$\beta$ to be zero and the scale factor $R'\sim t^{2/3}$ in homogeneous
flat dust universe,  $\tilde r_{AH}$ agrees with the expression in 
\cite{BR99,Bou99} for the apparent horizon. 

In order to calculate the entropy within the apparent horizon we have to
define the local  
entropy density first. From the standard big-bang cosmology \cite{KT} we
learnt that when a  
relativistic particle becomes non-relativistic and disappears, its entropy
is shared between the particle become thermal contact. Since 
photons and neutrinos never become non-relativistic, they share the
entropy of the universe. It is reasonable to suppose that the entropy of 
the universe is mainly produced before the dust-filled era and this 
result should also hold in the inhomogeneous cosmology. Using the first 
law of thermodynamics, for the dust-filled universe $(p=0)$ we 
have the local entropy density $s=\rho/T$, where $T$ is the temperature of
the universe and $\rho$ the radiational energy density given by $aT^4$.

Considering that in the expansion of the universe, the radiation always
has the property of black body and supposing that the number density 
of the photon is conserved, we also
have the relation $\dpf{h\nu_0}{kT_0}=\dpf{h\nu}{kT}$ in the inhomogeneous
background where $\nu_0=\nu (1+z)$. From the geodesic equation 
the expression for the redshift is 
\beq      \label{6}
1+z=\dpf{dt}{d\lambda}\vert_{\lambda}(\dpf{dt}{d\lambda}\vert_{\lambda=0
})^{-1} =R'\dpf{dr}{d\lambda}\vert_{\lambda}(R'_0\dpf{dr}{d\lambda}
\vert_{\lambda_0})^{-1}  
=\dpf{R'}{R'_0} 
\eeq
where assumed that for $r\rightarrow 0, R'=1$ \cite{Rib}. We obtain the
relation 
$TR'=T_0R'_0=Const.$, which coincides with that the homogeneous
case. Combining these 
considerations, the local entropy density in the inhomogeneous case can be
expressed as 
\beq    \label{7}
s(t,r)= a(T_0R'_0)^3\dpf{1}{R'^3(t,r)}=C\dpf{1}{R'^3(t,r)}
\eeq
where $C$ is a constant and $R'(t,r)$ has the form given in (\ref 2). 
The total entropy measured  
in the comoving space inside the apparent horizon is
\beq  \label{8}
S=\int^{r_{AH}}_0 s(t,r)4\pi R'R^2dr.
\eeq
For the homogeneous dust universe the local entropy density is
only a function of $t$ proportional to $a^{-3}(t)$ from the first
law. Eq(\ref 8) can 
thus reproduce the value $S=\dpf{4\pi}{3}\sigma r^3_{AH}$, where
$\sigma$ denotes the constant comoving entropy density.

With the method of calculating the total entropy of the inhomogeneous
parabolic model at hand, we state our proposal of a holographic principle 
in inhomogeneous cosmology in the 
spirit of \cite{BR99,Bou99}: the entropy inside the apparent horizon can 
never exceed the area of apparent horizon in Planck units. 

In order to get the entropy value and examine the holographic principle,
we need detailed expressions of $F(r), \beta(r)$. Two particular 
forms of these arbitrary functions that 
lead to fractal behavior in parabolic models have been found in
\cite{Rib}. They are 
\beq     \label{9}
{\rm Model \hspace{1.5mm}1}: 
   F=\alpha r^p \quad ,\quad
    \beta = \beta_0 +\eta_0 r^q
\eeq
\beq     \label{10}
{\rm Model \hspace{1.5mm}2}: 
    F=\alpha r^p \quad ,\quad
    \beta = \ln(e^{\beta_0}+\eta_1 r)
\eeq
where $\alpha\in[10^{-5}, 10^{-4}]$, $ p$  and $\beta_0\in [0.5, 4],
\eta_1\in[1000, 1300], q$ around 0.65 and $\eta_0$ around 50 are required
to obtain fractal solutions.

The starting point of the dust-filled universe is at $t_0=10^{12} s$
and the present time of the universe $t=15$ Gyr.
Considering the very large numbers appearing in the numerical calculations, we
adopt the units as those used in \cite{Rib}. We express distances in
$Gpc$, time unit  
in $3.26$ Gyr, mass unit (MU) as $2.09\times 10^{19}M_{\odot}$ and the
temperature as 
$1K=3.7\times 10^{-93} MU$ to keep $c=G=k=1$, ($k$ is the Boltzman
constant). 
Using the present temperature $T=2.7 K$, and the present size
$10^{28} cm$ \cite{KL99}, in our units the constant $C$ in (\ref 7) amounts to
$2.76\times 10^{86}$.

Now, we start to investigate in detail these (fractal) parabolic models.
Substituting (\ref 9) into (\ref 4), the proper apparent horizon can be 
gotten by solving the nonlinear equation
\beq        \label{11}
\alpha r_{AH}^p=3(t+\beta_0+\eta_0 r_{AH}^q).
\eeq
We found through analytical analysis that (\ref{11}) has no solutions when
$p<q=0.65$, which corresponds to having no apparent horizon in
that range for $p$. This is not so that surprising if we recall the fact that
not all homogeneous universe models have particle or event horizons. But
compared to the flat dust FLRW model, which always has apparent horizon,
this result indicates the difference between the general parabolic LTB
model and its homogeneous counterpart. For $p>q$, the proper apparent
horizon can be obtained by solving (\ref{11}) and the physical apparent
horizon is
\beq       \label{12}
\tilde{r}_{AH}=\dpf{1}{2}\alpha r_{AH}^p.
\eeq
We found that changes of $p$ and $\alpha$ change a lot the behavior of
the solution. Bigger $p$ or $\alpha$ leads to smaller results for $r_{AH}$.
The difference caused by different values of $\alpha$ for
big $p$ is smaller compared to that for small $p$. $\beta_0$ here does not
affect much the result.

The area of the apparent horizon in Planck units reads
\beq   \label{13}
A/l^2_p=4\pi b\tilde{r}_{AH}
\eeq
where $b=0.36\times 10^{121}$ in our units. With (\ref 9) and the obtained
value for  $r_{AH}$, the entropy inside the apparent horizon is
\begin{eqnarray}   
S&=&2.76\times 10^{86} 9\pi\times\label{14}\\
&&\int^{r_{AH}}_0\dpf{(t+\beta_0+\eta_0
r^q)^2}{[\dpf{p}{2}r^{-1}(t+\beta_0+\eta_0 r^q)+\eta_0 q r^{q-1}]^2}dr\; .
\nonumber
\end{eqnarray}
From the value of the constants in (\ref{13}) and (\ref{14}), one might
naively expect that the holographic principle always holds. However this
is not true.

\begin{center}
\setlength{\unitlength}{1.0mm}
\begin{picture}(75,37)(0,0)
\thicklines
\put(0,0){\resizebox{37\unitlength}{37\unitlength}%
{\includegraphics{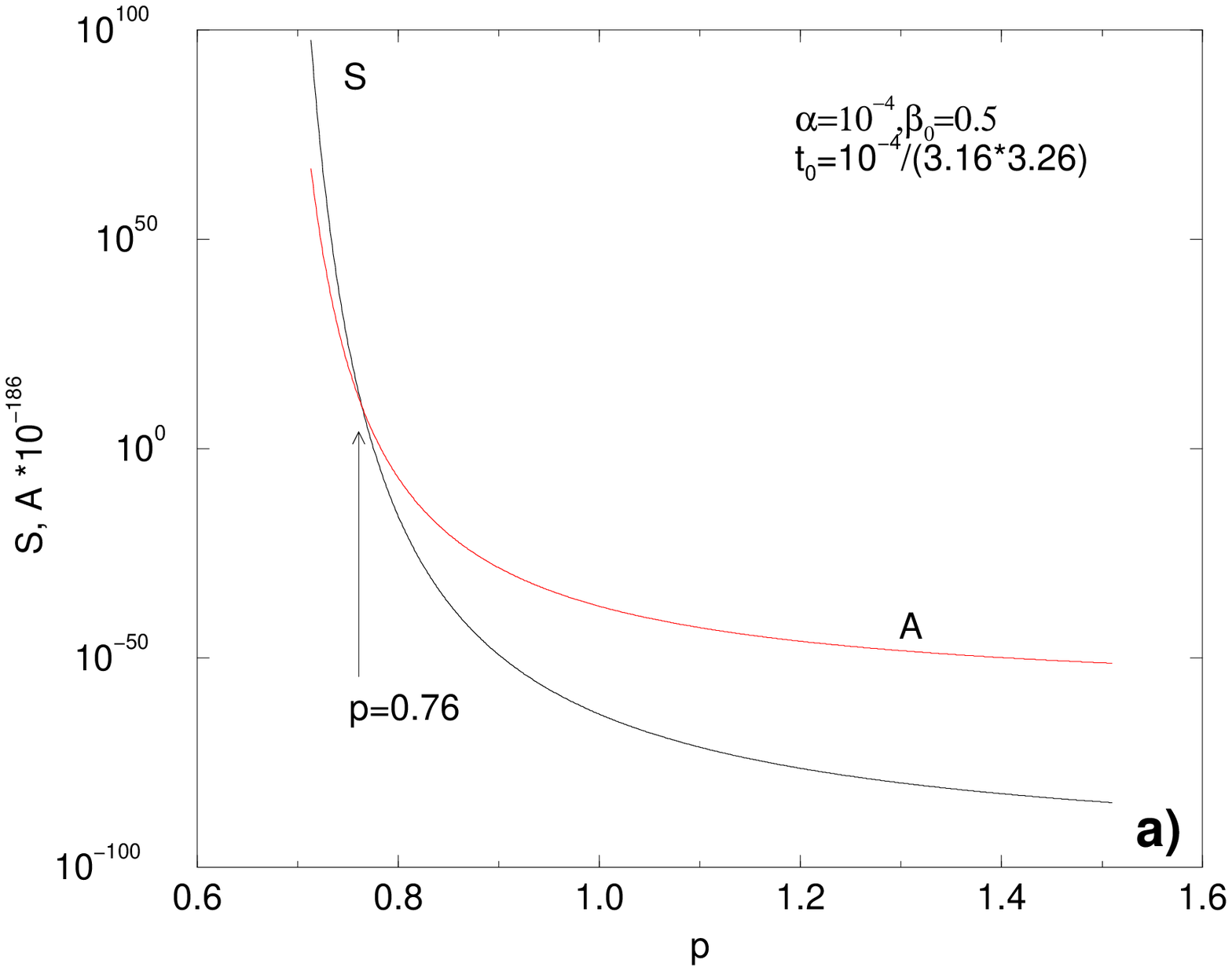}}}
\put(40,0){\resizebox{37\unitlength}{37\unitlength}%
{\includegraphics{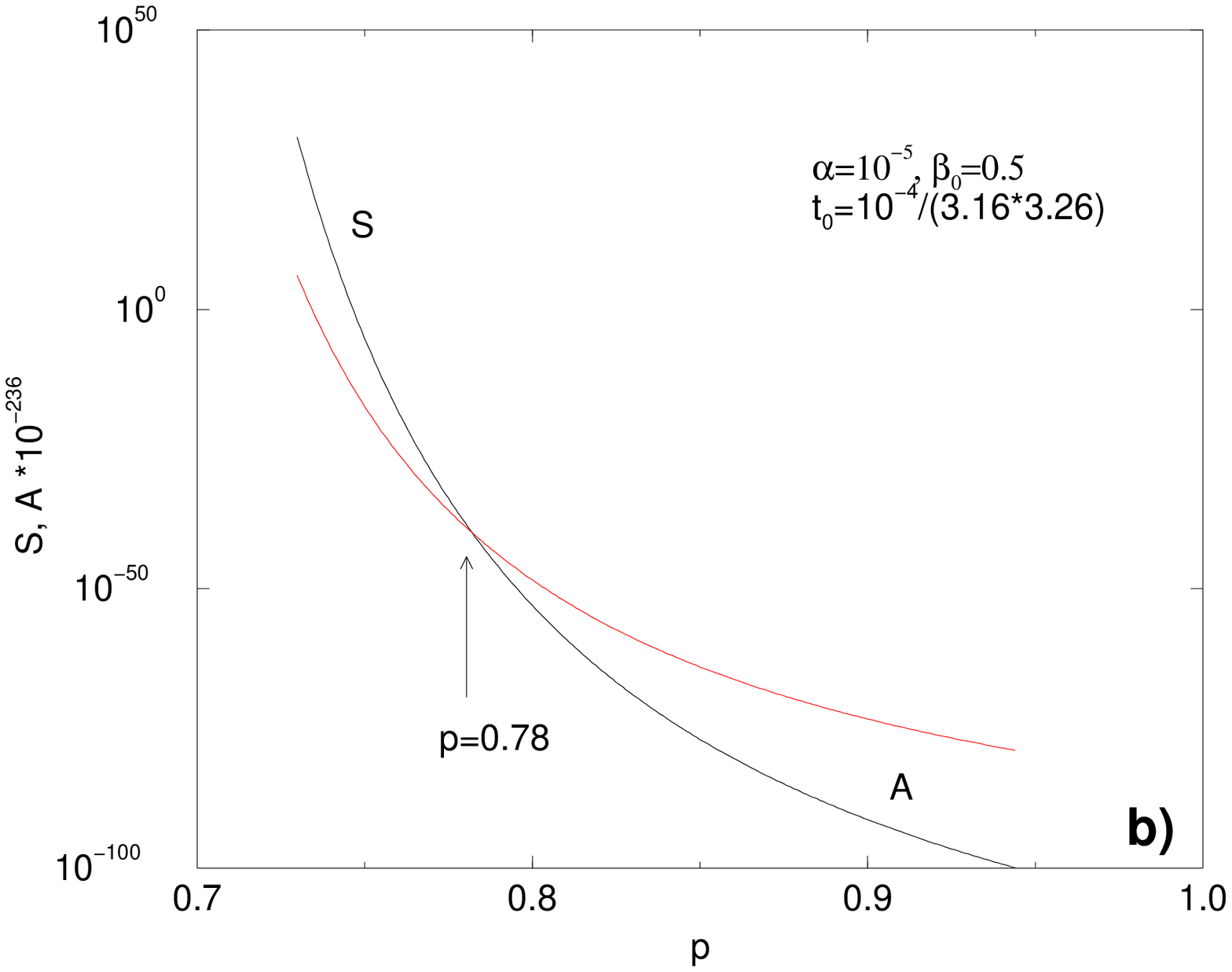}}}
\label{fig:12a}
\end{picture}
\parbox[t]{75mm}{Fig. 1: Relation between $S$ and $A$ with different
$p$ at the beginning of the dust-filled universe when $t_0=0.97
\times 10^{-5}$. }
\end{center}

Fig. 1 shows that at the beginning of the dust-filled era,
$t_0=0.97\times 10^{-5}$ in our units, when 
$\alpha=10^{-4}$ or $10^{-5}$, the holographic 
principle will be violated if $p<0.76$ or $p<0.78$, respectively. This
result does not change much for different values of $\beta_0$. 
The violation of the holographic principle here is really surprising 
because in homogeneous expanding universes, the holography has never 
been reported facing any challenge. This again shows the difference
between fractal parabolic models and its special homogeneous counterpart. 

We now face the question whether the holographic principle has to be
challenged or it can be used to select a physically acceptable model. 
We prefer the second, more constructive, alternative.

It is well believed that the entropy of the present observable universe is
of order $10^{90}$ \cite{KL99,Ven99}. This reasonable entropy value can be 
used as a standard to select models describing the real cosmos. 
\vskip 2cm

\begin{center}
\setlength{\unitlength}{1.0mm}
\begin{picture}(75,37)(0,0)
\thicklines
\put(0,0){\resizebox{37\unitlength}{37\unitlength}%
{\includegraphics{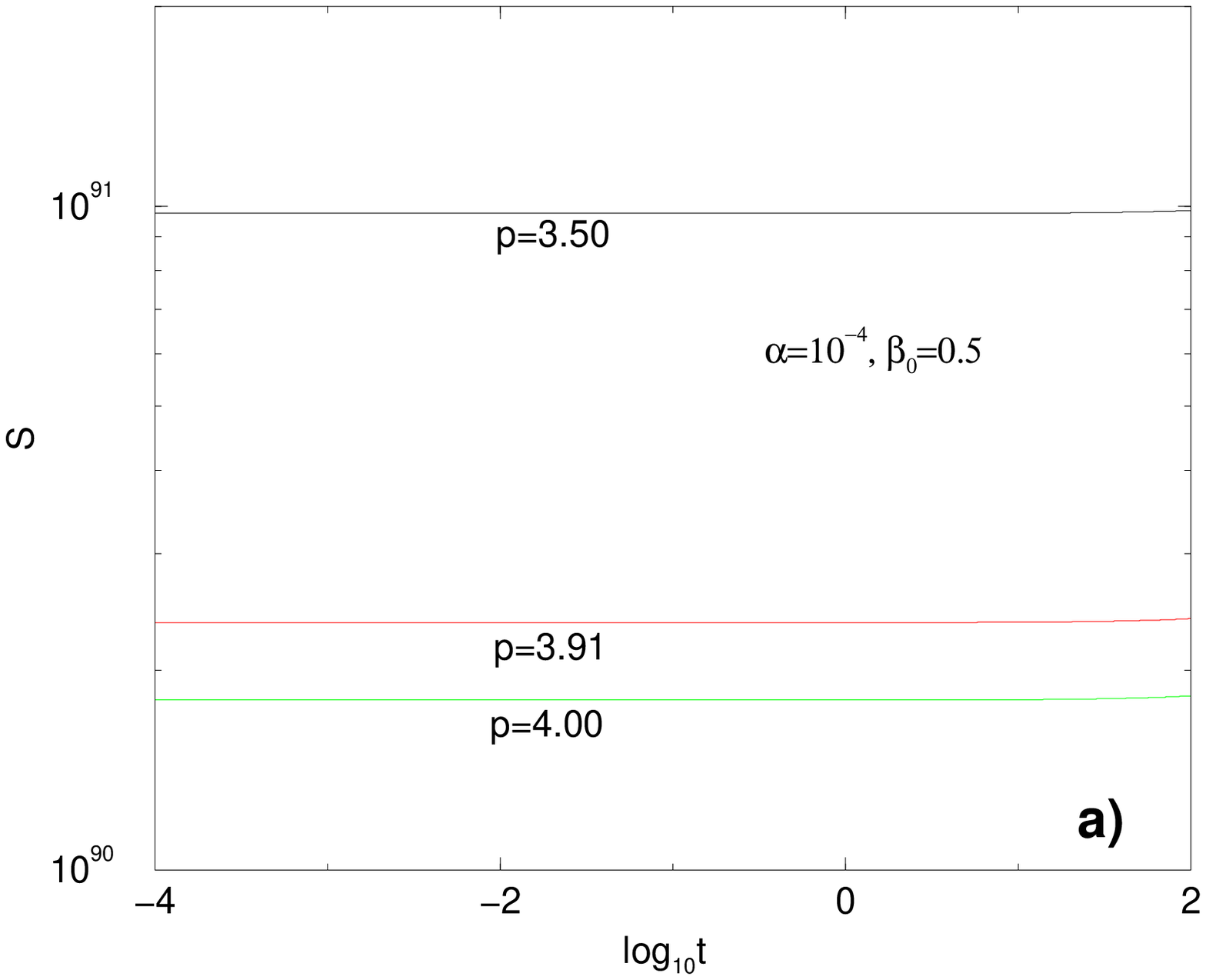}}}
\put(40,0){\resizebox{37\unitlength}{37\unitlength}%
{\includegraphics{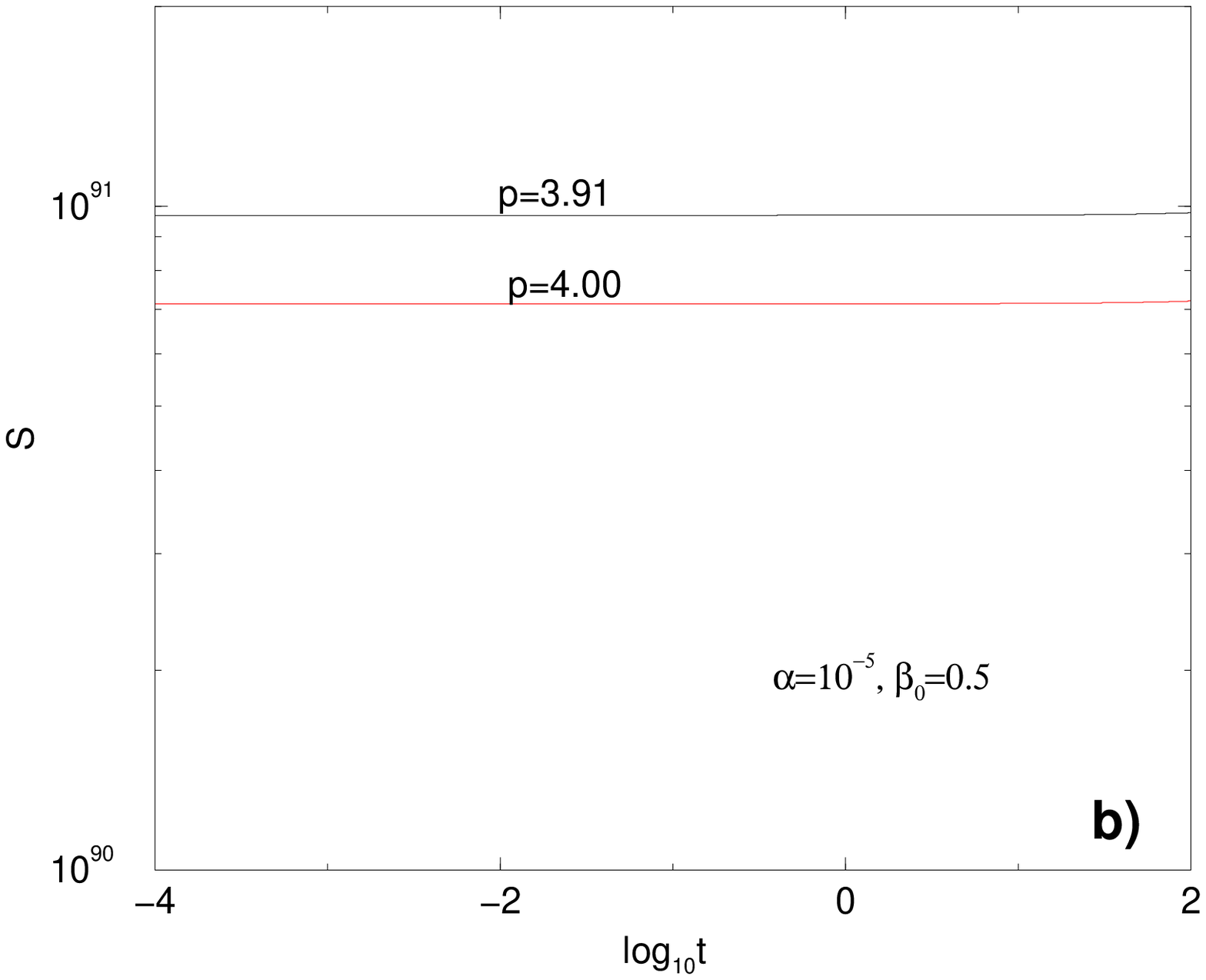}}}
\label{fig:12b}
\end{picture}
\parbox[t]{75mm}{Fig. 2: Inhomogeneous models which can accomodate
reasonable entropy to meet the present observable value. }
\end{center}

Fig.2(a) shows that for $\alpha=10^{-4}, 3.50\leq p\leq 4.0$, the entropy
values of the universe described by the fractal model is around $10^{90}$. 
(b) shows that for $\alpha=10^{-5}$, the range of $p$ changes to 
$3.91\leq p\leq 4.0$ to meet the required 
reasonable entropy. The influence of different $\beta_0$ is small. These
results show us how to characterize the constants of the model. 
It is worth noting that the range of $p$   
violating the holographic principle has been excluded here, which
corresponds to say that if these fractal models describe the real 
universe, they must satisfy the holographic principle. 

The dependence of the entropy value on constants of the  model shown in
Fig.2 is similar to that of $r_{AH}$. Bigger values of $p$ leads to 
smaller values of entropy, and 
for the same $p$, bigger $\alpha$ brings smaller entropy. 

Now we extend our discussion to Model 2. Using (\ref{10}), the proper apparent
horizon can be got from
\beq  \label{15}
\alpha r_{AH}^p=3[t+\ln(e^{\beta_0}+\eta_1 r_{AH})].
\eeq
In contrast to Model 1, apparent horizon can be found for all constants
displaying fractal behavior. $p$ and $\alpha$ play the same crucial 
role to influence the result of $r_{AH}$. While the influence of 
$\beta_0, \eta_1$ is not important. The area of the 
apparent horizon in Planck units is expressed by (\ref{13}) and the total
entropy inside the apparent horizon is
\begin{eqnarray}        
S&=&2.76\times 10^{86} 9\pi\times\label{16}\\
&&\int^{r_{AH}}_0\dpf{[t+\ln(e^{\beta_0}+\eta_1
r)]^2}{[\dpf{p}{2}(t+\ln(e^{\beta_0}+\eta_1
r))r^{-1}+\dpf{\eta_1}{e^{\beta_0}+\eta_1 
r}]^2}dr. \nonumber
\end{eqnarray}
We found that the holographic principle always holds for this model. However
for small values of $p$, the result for the entropy is again too big to
meet the requirement of describing an observable realistic universe. In 
order to make the inhomogeneous model reasonable to describe the 
realistic universe, we need the observed entropy value as a criteria to 
choose reasonable constants in the model.

\begin{center}
\setlength{\unitlength}{1.0mm}
\begin{picture}(75,37)(0,0)
\thicklines
\put(0,0){\resizebox{37\unitlength}{37\unitlength}%
{\includegraphics{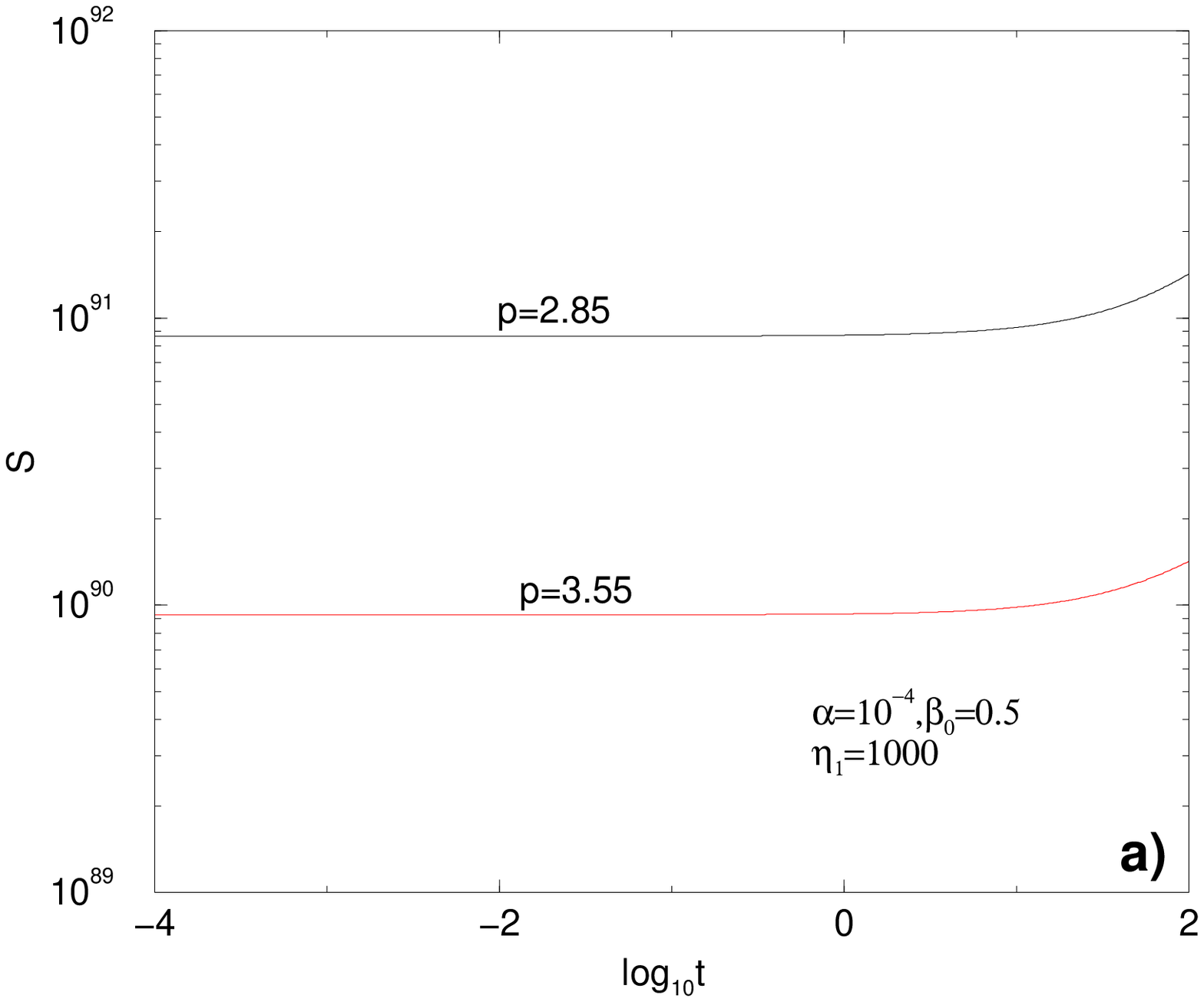}}}
\put(40,0){\resizebox{37\unitlength}{37\unitlength}%
{\includegraphics{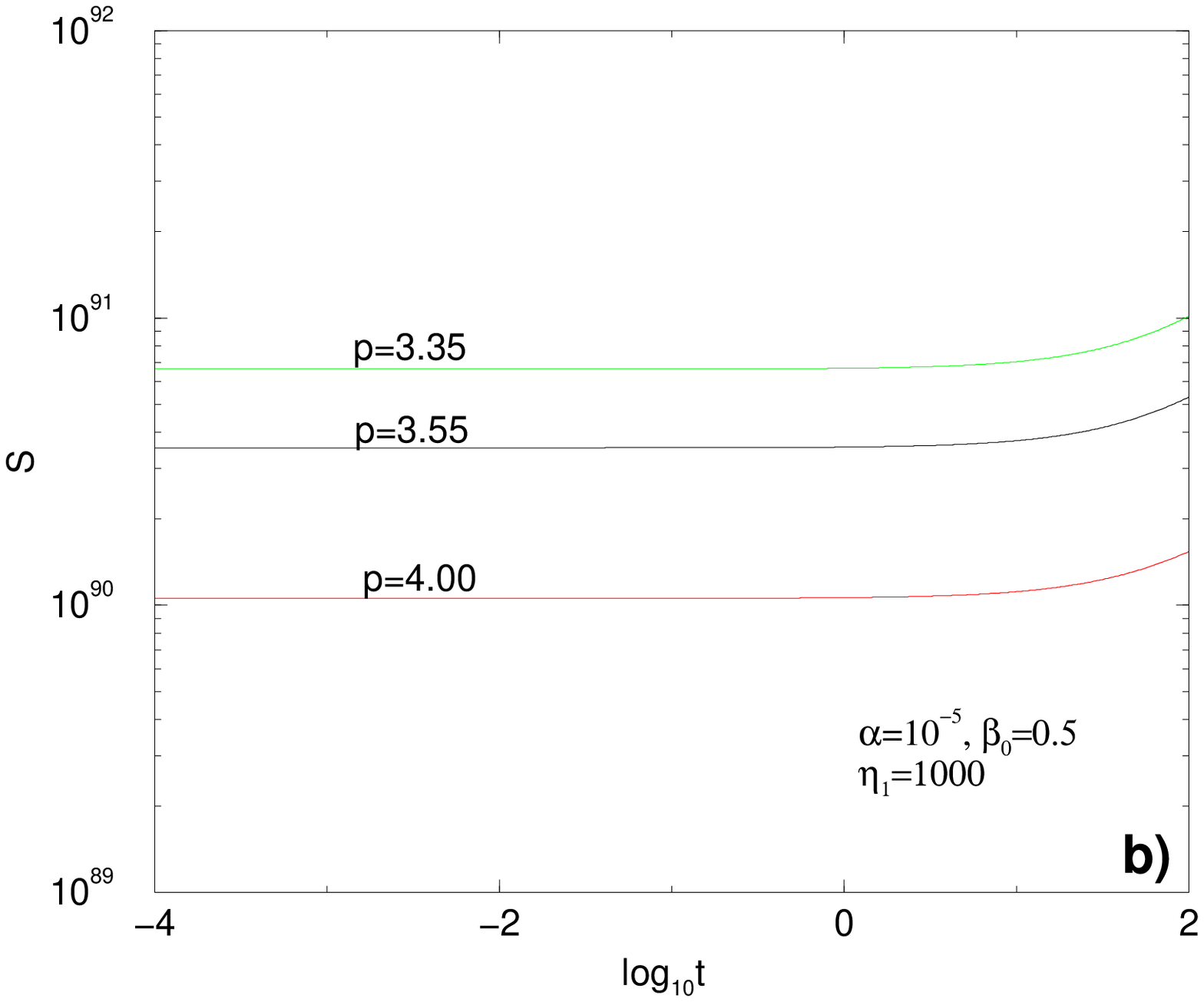}}}
\label{fig:12c}
\end{picture}
\end{center}
\begin{center}
\setlength{\unitlength}{1.0mm}
\begin{picture}(75,37)(0,0)
\thicklines
\put(0,0){\resizebox{37\unitlength}{37\unitlength}%
{\includegraphics{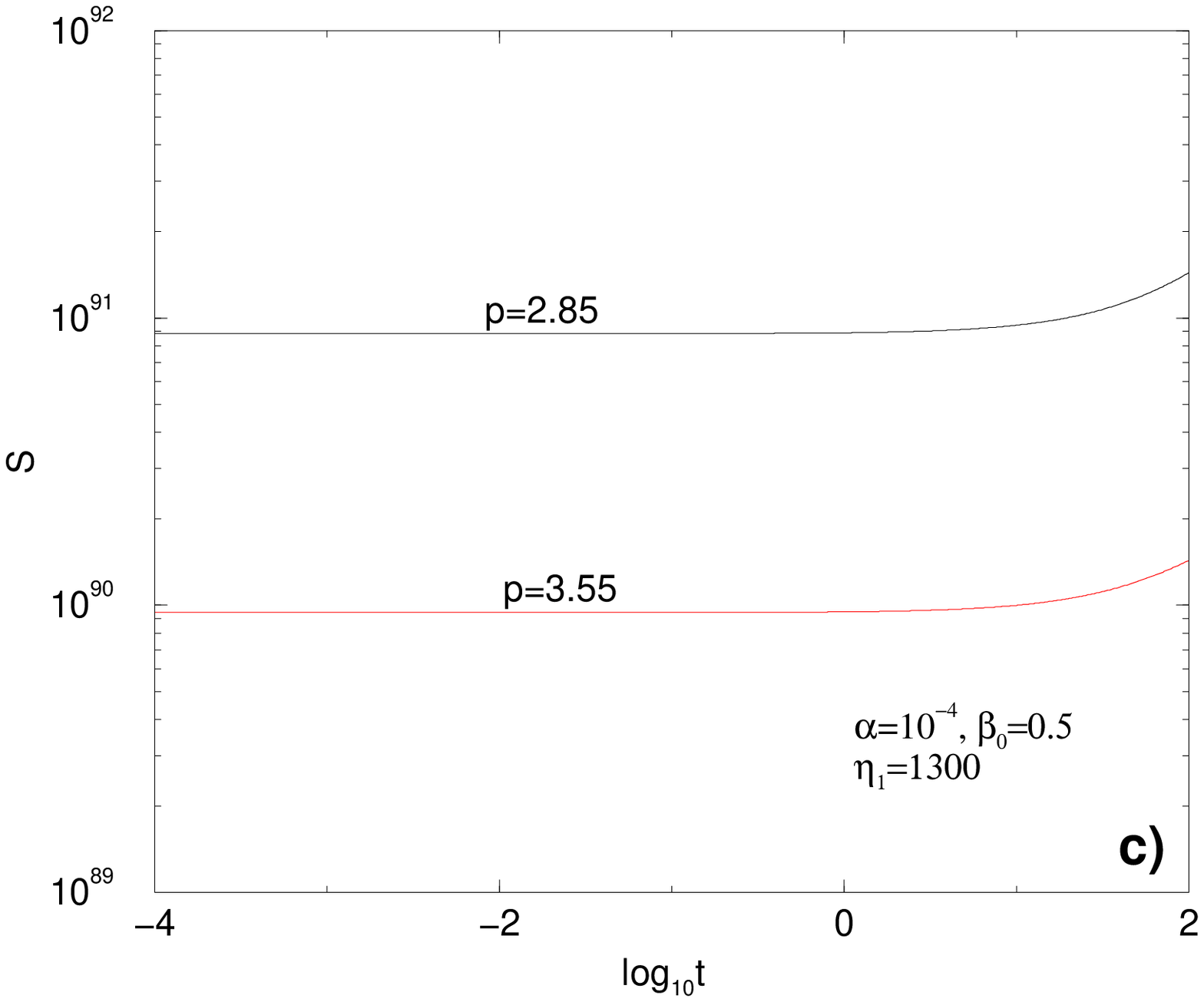}}}
\put(40,0){\resizebox{37\unitlength}{37\unitlength}%
{\includegraphics{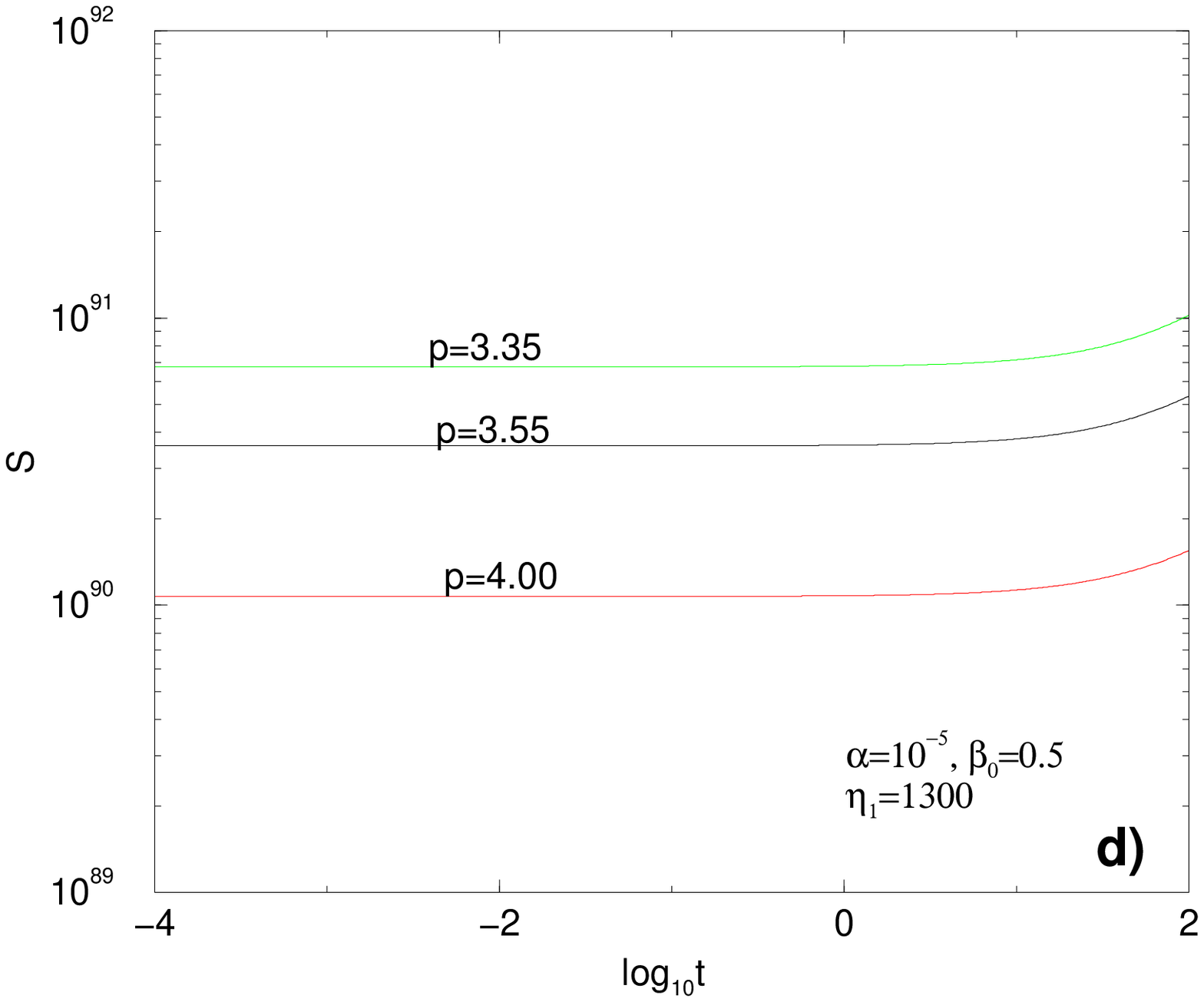}}}
\label{fig:12d}
\end{picture}
\parbox[t]{75mm}{Fig. 3: Choosing parameters in order to meet the 
entropy value in the present observable universe.}
\end{center}

Fig.3 presents some regions of constants which can display a reasonable
entropy value. (a) and (b) show that for $\alpha=10^{-4}, 2.85\leq 
p\leq 3.55$; and for $\alpha=10^{-5}, 3.35\leq p\leq 4.0$, the total 
entropy value at the present time obtained in Model 2 can meet the 
estimative value $S\sim 10^{90}$. These regions of
constants are required to delineate the real cosmos by this fractal
model. From Fig.3 we find that influence of $p$ and $\alpha$ on the 
entropy value is the same as that discussed in Model 1. Bigger $p, 
\alpha$ corresponds to smaller $S$. Comparing (a) and (b), (c) and (d), 
we learn that different $\eta_1$ does not change a lot of the final 
result. This behavior also holds for $\beta_0$.

In summary, considering properties of spherical symmetry and not relating
to either initial or final moment for inhomogeneous dust universes, 
we introduced a simple holographic principle by asserting that all 
information about physical processes in the real cosmological setting 
can be stored on the surface of the apparent horizon. 
Investigating fractal parabolic models with the holographic principle, we
found that the violation of the holographic principle appears in some 
fractal parabolic models, what has never been observed in any special 
flat homogeneous universe. In order to describe 
the real cosmos, we refined the fractal parabolic models by restricting
constants 
choosing regions to get reasonable entropy values of the present
observable universe. We 
found that the realistic models for an inhomogeneous universe satisfy the 
holographic principle. The slowly increasing of the entropy value with
evolution of time 
in the inhomogeneous dust universe supports the behavior illustrated in
homogeneous 
cosmology \cite{Ven99}, which shows that the entropy in the universe is
mainly created before 
the dust-filled era.

ACKNOWLEDGMENTS:
This work was partially supported by Fundac\~ao de Amparo \`a Pesquisa 
do Estado de S\~ao Paulo (FAPESP) and Conselho Nacional de Desenvolvimento
Cient\'{\i}fico e Tecnol\'{o}gico (CNPQ). B. Wang would  like to acknowledge 
the support given by Shanghai Science and Technology Commission.

\end{document}